\documentclass{iopart}
\usepackage{iopams}
\usepackage{graphicx}
\usepackage{color}
\usepackage{ulem}
\usepackage{listings}
\usepackage{subfig}

\begin{document}
\title[New Electrostatic Mirror Actuators for High-Precision Interferometry]{New Design of Electrostatic Mirror Actuators  
for Application in High-Precision Interferometry}
\author{H\,Wittel$^{1}$, S\,Hild$^{2}$ G\,Bergmann$^{1}$, K\,Danzmann$^{1}$ and K\,A\,Strain$^{2}$}
\ead{Holger.Wittel@aei.mpg.de}
\vskip 1mm
\address{$^{1}$\,Max--Planck--Institute for Gravitational Physics and Leibniz University of Hannover, D-30167 Hannover, Germany}
\address{$^{2}$\,SUPA, School of Physics and Astronomy, The University of Glasgow, Glasgow, G12\,8QQ, UK}

\begin{abstract}
We describe a new geometry for electrostatic actuators to be used  in sensitive laser interferometers. The arrangement consists of two plates at the sides of the mirror (test mass), and therefore does not reduce its clear aperture as a conventional electrostatic drive (ESD) would do. Using the sample case of the AEI-10m prototype interferometer, we investigate the actuation range and influences of relative
misalignment of the ESD plates in respect to the test mass. We find that in the case of the AEI-10\,m prototype interferometer, this new kind of ESD could provide a range of 0.28\,$\mu$m when operated at a voltage of 1\,kV. In addition, the geometry  presented is shown to provide a reduction factor of about 100 in the magnitude of actuator motion coupling to test mass displacement. We show that therefore in the specific case of the AEI-10m interferometer it is possible to mount the ESD actuators  directly on the optical table, without spoiling the seismic isolation performance of the triple stage suspension of the main test masses.
\end{abstract}


\section{Motivation}
\begin{figure}[htbp]
	\centering
		\includegraphics[width=1.00\textwidth]{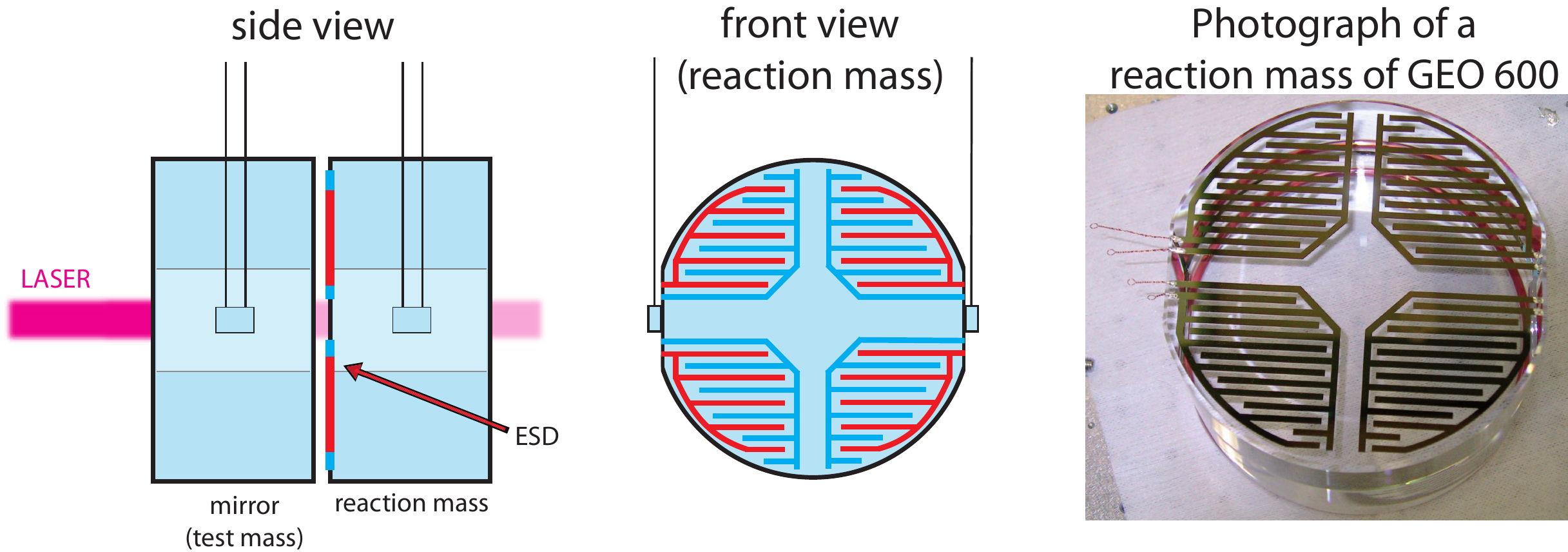}
	\caption{\\\textbf{Left and center:} Schematic drawing of a conventional ESD setup. The shown ESD setup consists of four independent ESD comb-shaped electrode pairs on a single reaction mass. The anodes are colored in red, the cathodes are colored blue.
	The center of the reaction mass is kept free so that a laser
	beam can pass without significant clipping.
	\\
	\textbf{Right:} Photograph of a reaction mass of GEO\,600}
	\label{fig:ESD_traditional}
\end{figure}

Interferometric gravitational wave detectors, such as GEO\,600 \cite{GEO}, Advanced LIGO (aLIGO) \cite{aLIGOoverwiew}, Advanced Virgo \cite{avirgo} and KAGRA \cite{KAGRA} are large laser interferometers with the mirrors/test masses hung at the bottom of multi-stage pendulum chains. For the operation of these detectors, it is necessary to have low-noise, contact-free actuators for controlling the longitudinal and angular degrees of freedom of the mirrors. This is traditionally done with either magnet-coil  actuators or electrostatic drives (ESDs). GEO\,600 has operated since 2001 with ESDs as the main longitudinal actuators for controlling the differential arm length. Based on this experience, ESDs are now employed in aLIGO, using a very similar configuration to the original GEO\,600 design.
The GEO/aLIGO ESDs are designed in the form of a metallic comb structure that has been coated onto a reaction mass and is located a few millimeters behind the mirror. Schematic drawings and a photograph of this conventional ESD setup are shown in Figure \ref{fig:ESD_traditional}.
The ESD on the reaction mass needs its own seismic isolation, to avoid the coupling of ground motion to the seismically isolated test mass. Therefore, it is also hung as the lowest stage of a multi stage pendulum, which again requires its own sensors and actuators for alignment and damping of the reaction mass. Furthermore, this conventional type of ESD reduces the possible free aperture in transmission. This may be problematic for experiments which require the largest possible free aperture, such as the planned AEI-10m prototype interferometer \cite{graef}\cite{gossler} or the speedmeter proof of principle experiment in Glasgow \cite{speedm1}\cite{speedmGraef}.

With the AEI-10m interferometer in mind, we investigate a new ESD configuration, featuring a simpler geometry, that uses only two plates at the sides or at top and bottom of the mirror, pictured in Figure \ref{fig:setup}. Since the force that may be obtained per applied voltage of this setup is smaller than for the conventional ESD configuration, this type of ESD is mainly suited for the case of light ($\leq$ 100\,g) mirrors such as in the AEI-10m prototype interferometer or in the Glasgow speed meter proof of principle experiment.

\section{Basic Principle}
\begin{figure}
	\centering
		\includegraphics[width=1.00\textwidth]{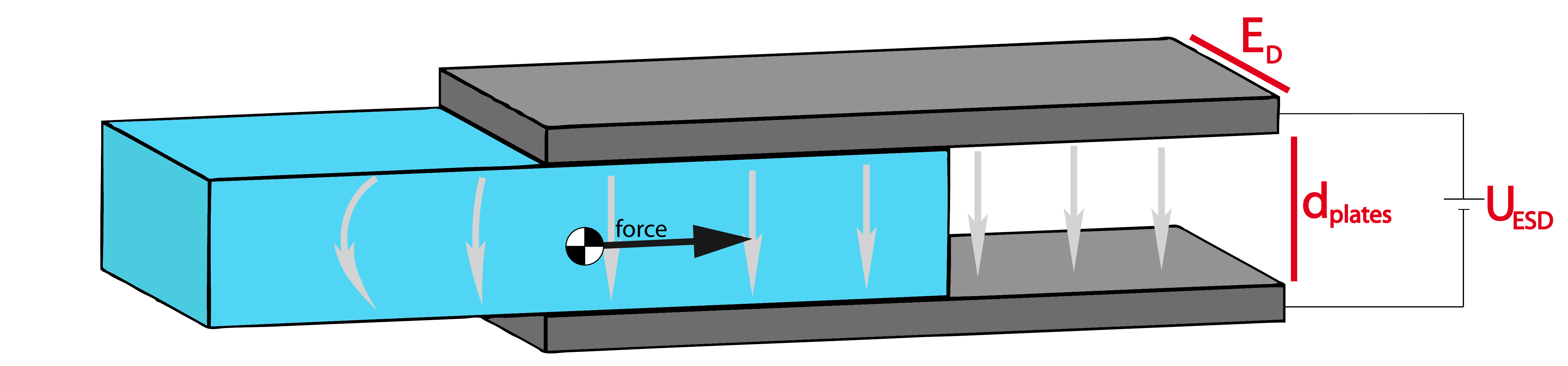}
		\caption{Illustration of the simple model. A slab of a dielectric material between two parallel capacitor plates. The electric field is pictured as gray arrows.}
	\label{fig:simplemodel}
\end{figure}

The working principle of ESDs is that an inhomogeneous electric field is built up in a dielectric medium (i.e.~the mirror). If the mirror is not centered longitudinally between the plates, it will be subject to a force that pulls it towards the center of the plates. In order to analyze our new ESD design, we first turn to a simplified analytic model, before we use finite elements (FE) simulations later. Our simplified model shall be the case of a dielectric slab inserted into a parallel-plate capacitor (as in \cite{dielectric}), which is pictured in Figure \ref{fig:simplemodel}.
Calculating the force for this simple model is a standard problem in electrodynamics.
The force is given by \cite{dielectric}:
\begin{equation}
	F=(\epsilon-\epsilon_0)\cdot E_D \cdot U_{ESD}^2/ 2d_{\mathrm{plates}},\\
	\label{simplecalc}
\end{equation}
where $E_D$ is the plate depth, $U_{ESD}$ is the voltage across the plates and $d_{\mathrm{plates}}$ is the plate separation. $\epsilon_0$ and $\epsilon_r$ are dielectric constants (see also table \ref{tab:material} for the values and symbols used in this article). This formula is only valid under certain assumptions:
\begin{enumerate}
\item There is no gap between the dielectric and the plates.
\item One end of the dielectric material is in a homogeneous electric field between the plates.
\item The other end of the dielectric is far outside of the plates, it does not 'see' the electric field of the plates.
\end{enumerate}

We can use the formula for the simplified case to approximate the order of magnitude of the force that the ESD will provide. The exact strength of the force must be determined by finite element methods, since none of the assumptions mentioned above is exactly fulfilled. 

To estimate the force that the ESD can provide using the simplified formula, first we compute an effective dielectric constant $\epsilon_{r_{\mathrm{eff}}}$ for a mirror between the two plates, since it does not fill the space between the ESD plates completely. We determine a 'fill factor' of $A_{\mathrm{mirror}}/A_{\mathrm{plates}}=\pi r_{\mathrm{mirror}}^2 / E_D\cdot d_{\mathrm{plates}}$ = 0.54. Now we can multiply the fill factor with $\epsilon_r$ and get an effective $\epsilon_{r_{\mathrm{eff}}}=2$. With the values given in Table~\ref{tab:material} and an $\epsilon_{r_{\mathrm{eff}}}$=2, equation~\ref{simplecalc} gives a position independent force of the order of some $\mu$N.

\section{Quantitative Analysis using FEM}
\begin{figure}
\centering
		\includegraphics[width=1\textwidth]{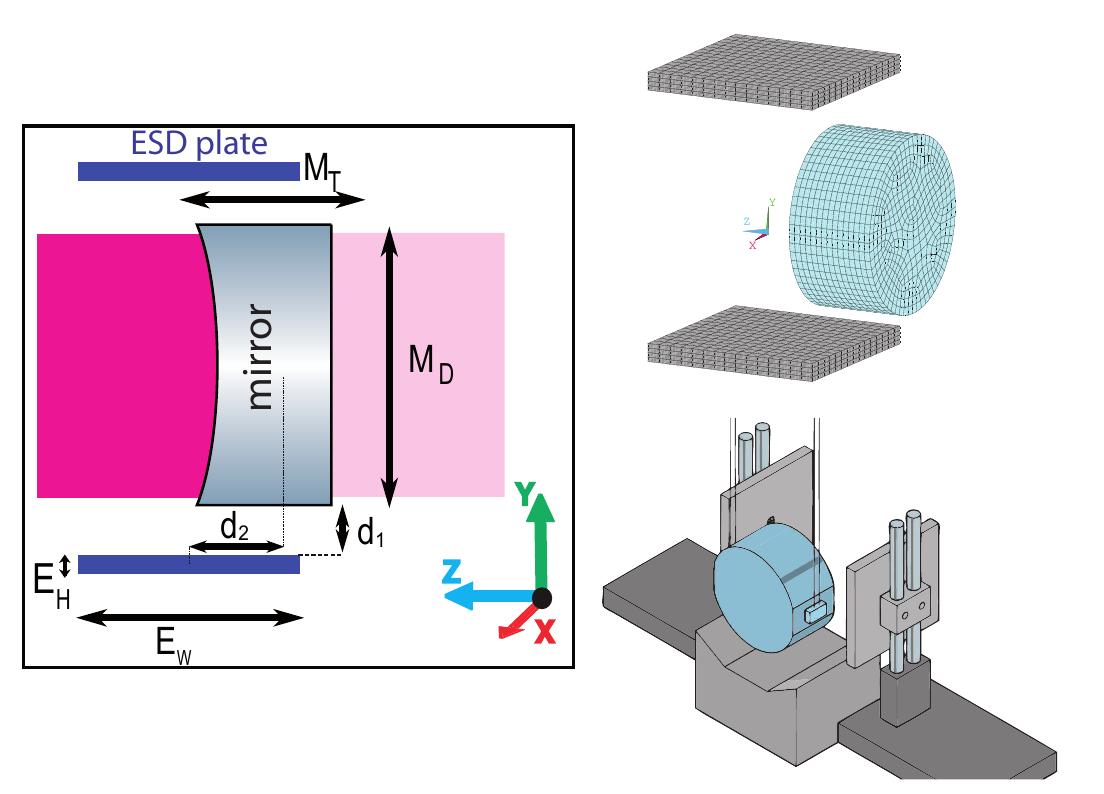}
	\caption{\\ \textbf{Left:} Basic geometrical setup of the ESD plates with respect to the main test masses/mirror. Please note that this sketch can be either side-view or top-view, depending on the actual installation. The probing laser beam is shown in purple while the beam transmitted through the mirror is painted with a lighter color.	The dimensions M$_{\rm T}$, M$_{\rm D}$, E$_{\rm H}$, E$_{\rm W}$, and d$_{1,2}$ are defined in table \ref{tab:material}.  \\\textbf{Top right:} View of the FE model that was used in this article.\\\textbf{Bottom right:} A mock up drawing of what a potential installation might look like.}
	\label{fig:setup}
\end{figure}
\begin{figure}[htbp]
\centering
\includegraphics [width=1\textwidth] {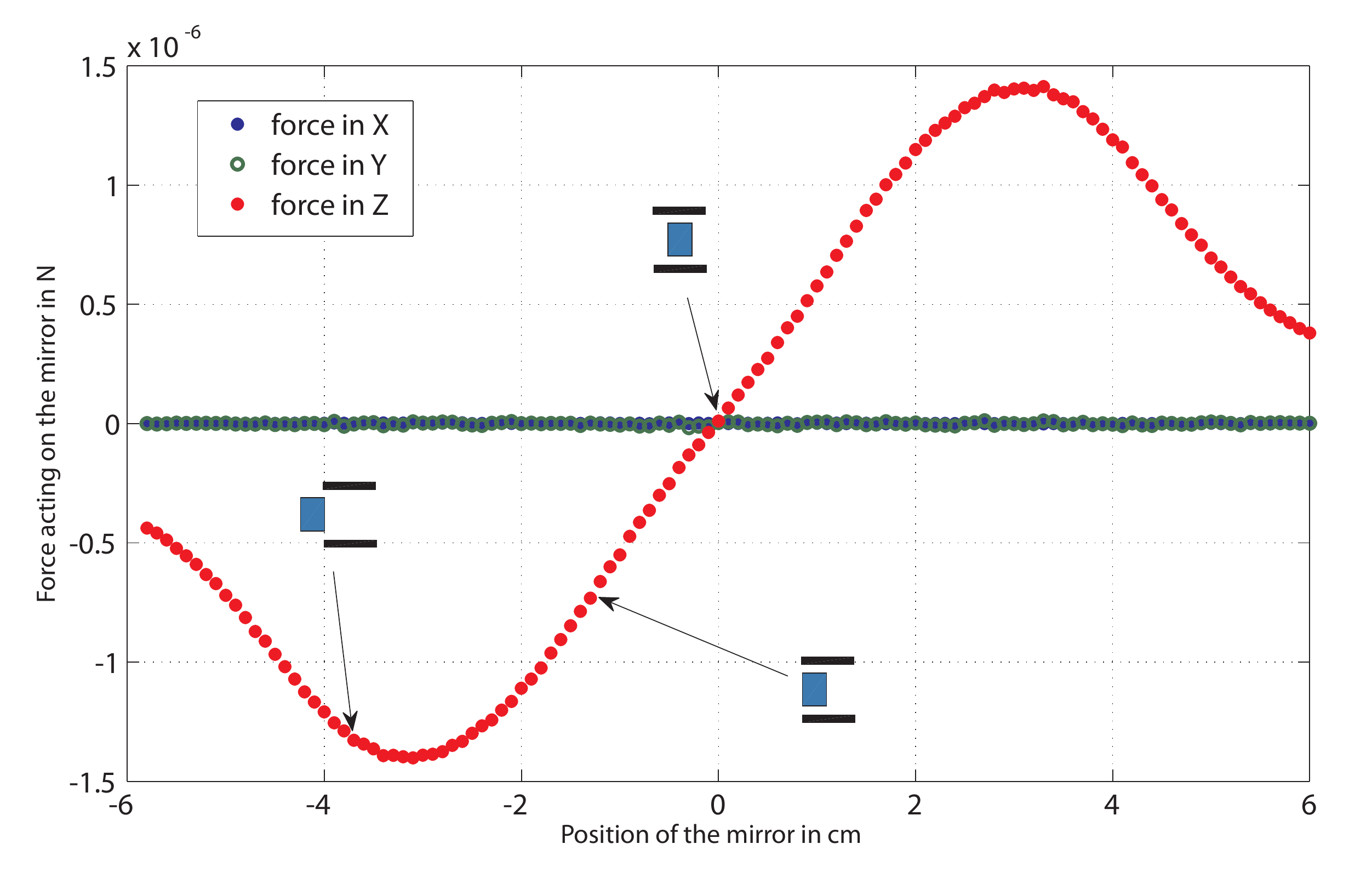}
\caption{Longitudinal force on the mirror, versus relative longitudinal position of mirror and ESD. The x-axis shows the relative distance between the mirror center and the center of the capacitor plates.}
\label{fig:force-pos}
\end{figure}

\begin{table}[h]
\begin{center}
\begin{tabular}{|r|c|c|}
\hline
Mirror diameter &M$_{\rm D}$ & 4.90\,cm \\
Mirror thickness &M$_{\rm T}$ &2.45\,cm \\
Mirror material & - &fused silica\\
Mirror dielectric constant &$\epsilon_{r}$ &3.7$\epsilon_{0}$\\
Mirror mass &m &102 \,g \\
Pendulum length & l & 20\,cm\\
Lateral distance mirror-ESD  &d$_{1}$ &1.05\,cm \\
Long. relative position mirror-ESD  &d$_{2}$ &3.075\,cm \\
Applied voltage &U$_{\rm ESD}$ &1000\,V\\
ESD dimensions (one plate) &E$_W$xE$_H$xE$_D$ &5\,cm x 0.5\,cm x 5\,cm \\
ESD plate separation &$d_{\mathrm{plates}}$& 7\,cm \\
Number of nodes & $n_{\mathrm{nodes}}$ &221944\\
\hline
\end{tabular}
\caption{Properties used for the FE simulations presented in this document.
 \label{tab:material}}
\end{center}
\end{table}
Unless noted otherwise, the values given in table~\ref{tab:material} are used for the simulations in this article. The origin of the coordinate system is located in the center between the ESD plates. The Z-axis points from the flat mirror surface to the center between the plates (sometimes called 'longitudinal direction' or 'beam direction'). The Y-axis points towards one of the plates and the X-axis is perpendicular to the Y and Z-axes. Figure \ref{fig:setup} shows the geometry of the new ESD setup.
The force that the ESD applies on the mirror in dependence of the relative longitudinal position between mirror and ESD is plotted in Figure ~\ref{fig:force-pos}.
From this dependency we also find the best operating distance between mirror and ESD: the maximum and to first order flat force of 1.4\,$\mu$N appears when the mirror is shifted by 3.075\,cm relative to the ESD plates. We choose this position as our potential operating point. 
From the force the magnitude of the ESD range can be deduced. We assume that the mirror is suspended as a pendulum of the length $l$. The full actuation range $x$ is reached when the pendulum back action force cancels the ESD-force:
\begin{equation}
	x=-F_{\mathrm{operating}}\cdot l/mg \approx 0.28 \,\mu $m$ $ (at 1\,kV)$.
	\label{eqn}
\end{equation}
From the equations \ref{eqn} and \ref{simplecalc} it also follows that the range is inversely proportional to mirror mass and plate separation. That is the reason why this new kind of ESDs is only suited for the case of small mirrors. 
\begin{equation}
	x\propto 1/(m\cdot d_{\mathrm{plates}})
	\label{eqn2}
\end{equation}

\begin{figure}[!ht]
    \subfloat[Voltage change \label{fig:volt-force}]{%
      \includegraphics [width=0.50\textwidth] {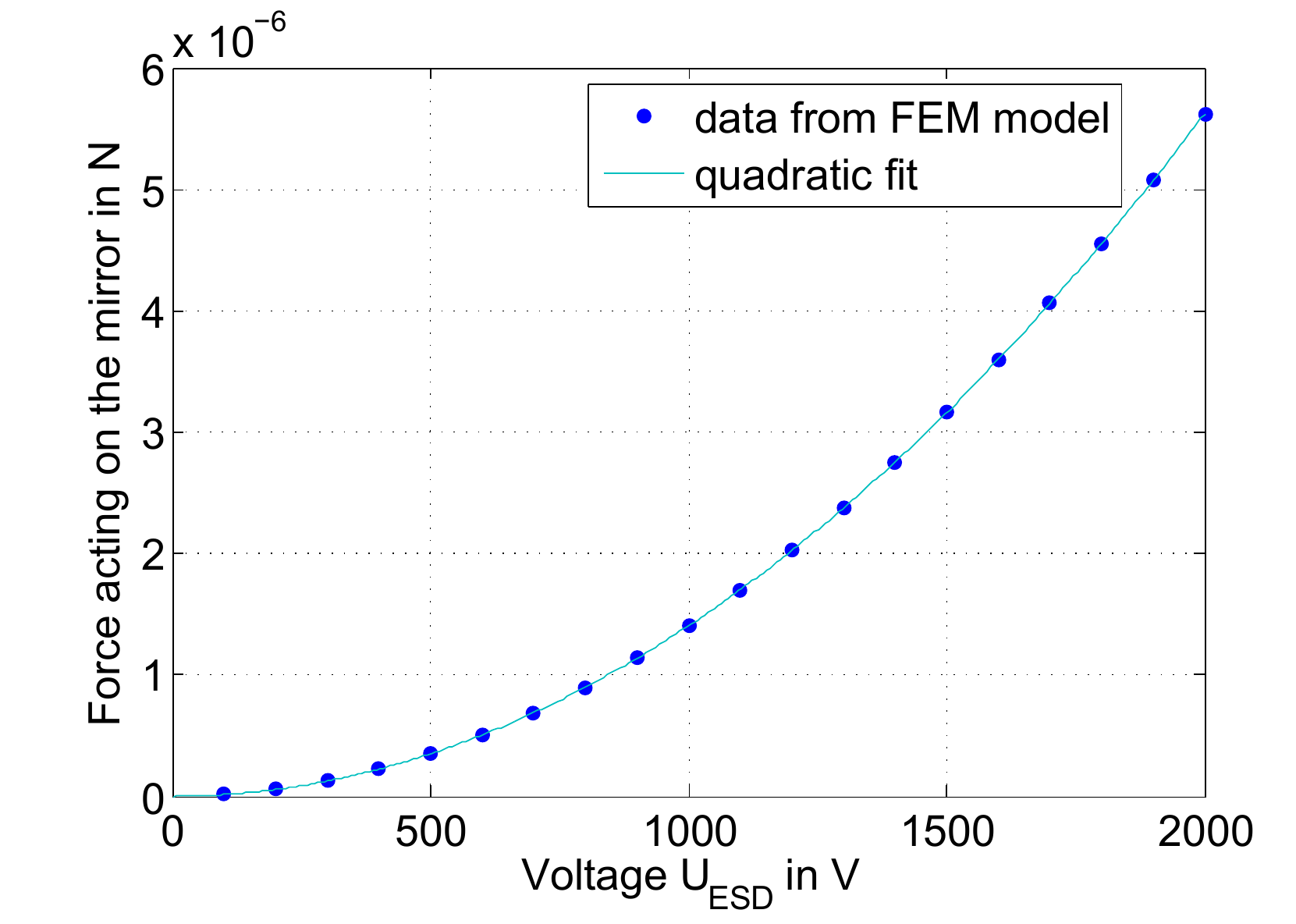}
    }
    \hfill
    \subfloat[Plate separation \label{fig:sep_force28-Sep-2011}]{%
      \includegraphics [width=0.50\textwidth] {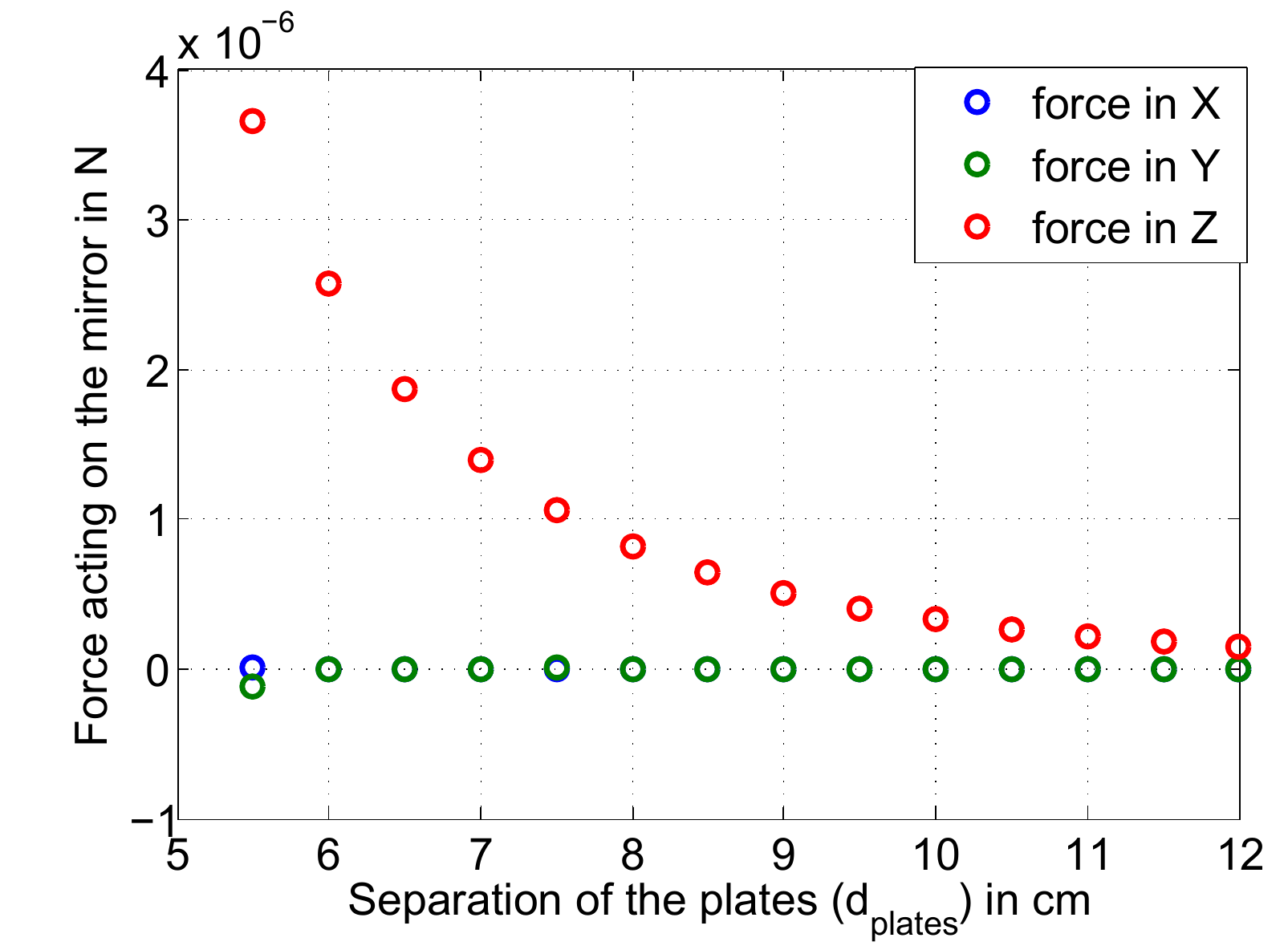}
    }
    \caption{Figure (a) shows the force on the mirror for different voltages. The relative position of mirror and ESD is the position with the optimal force. (b) shows how the force depends on the plate separation (assuming that the mirror is at the ideal operating position)}
  \end{figure}

Figure \ref{fig:volt-force} shows that the force that can be obtained scales quadratically with the applied voltage, when the mirror is in the operating position, while Figure~\ref{fig:sep_force28-Sep-2011} shows how the force depends on the plate separation.

\section{Requirements \& Noise}
\begin{figure}%
\centering
\includegraphics [width=1\textwidth] {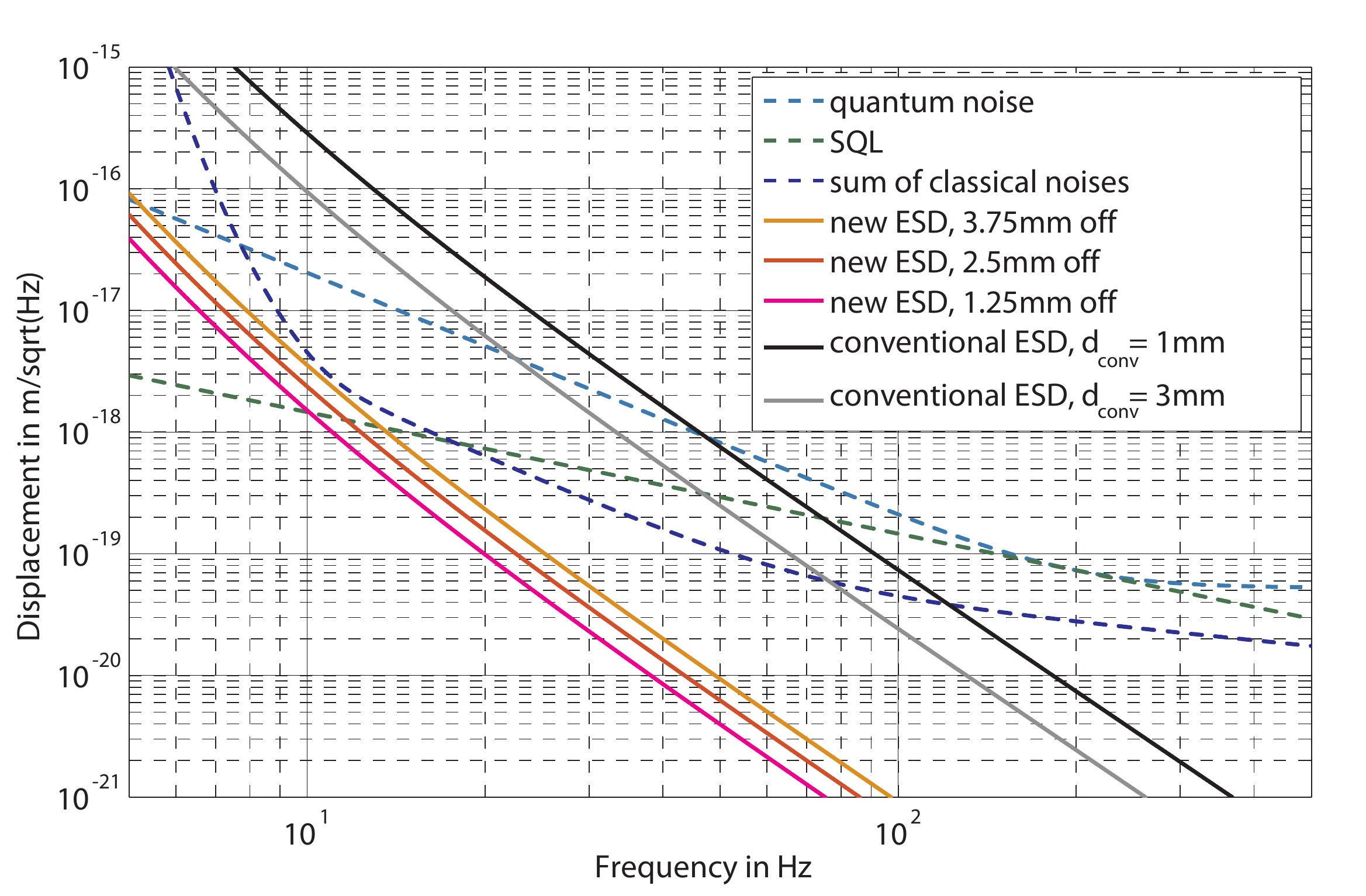}
\caption{Comparison of seismic noise coupling 
in the AEI-10m prototype interferometer for ESDs with the  
conventional design and our novel ESD configuration. 
The reddish traces indicate projections of how much table 
movement would couple into the interferometer if the novel ESD configuration is used, 
but the plates are not located ideally. The dashed lines 
indicate the sum of all classical noise, quantum noise and the Standard Quantum Limit (SQL) of 
  the planned AEI-10m sub-SQL interferometer. We have also included
  a projection of the seismic noise coupling if conventional ESDs
  mounted on the tables would be used, where d indicates the 
  distance between the mirror and the ESDs (greyish traces). 
  As one can see, the seismic noise coupling of our novel 
  ESD design is lower by about a factor 100 compared 
  to a conventional ESD design.}%
\label{fig:force_gradient2}%
\end{figure}

As mentioned above, the force on the mirror (provided by the new ESD configuration) is independent of the mirror position, as long as mirror and ESD are positioned correctly. Together with the performance of the seismically isolated tables in the AEI-10m prototype interferometer \cite{wanner} it is possible to mount the new ESD directly to the table without additional seismic isolation. Figure~\ref{fig:force_gradient2} shows the expected coupling of seismic noise through the new ESDs for non-ideal positioning of the mirror in respect to the ESD plates. Even with the ESDs 3.75\,mm off from the ideal longitudinal position, the noise contribution in the AEI-10m interferometer would still be lower than the sum of all other classical noise sources at all frequencies.
     
In order allow a direct comparison of the seismic coupling of a conventional ESD design and our new ESD design, we included the expected noise coupling from a conventional ESD design in Figure~\ref{fig:force_gradient2} (see grey traces). For this analysis we assumed the force on the mirror for a conventional ESD design, which is given by \cite{GroteDr}:
\begin{equation}
F = a \cdot \epsilon \cdot \epsilon_r \cdot  U_{ESD}^2 \cdot d_{\mathrm{conv}}^{1.5},
\end{equation}
 where $d_{\mathrm{conv}}$ is the separation between the mirror and the ESD
 and $a$ is a geometry factor that depends on the actual shape of the
 electrode pattern, and the dimensions of the mirror and the ESD. The strong 
 scaling of the force with $d_{\mathrm{conv}}$, explains the large coupling of seismic from the ESD to the test mass. This can be seen in Figure~\ref{fig:force_gradient2}, where we compare a conventional ESD to the new ESD design (both actuators provide the same longitudinal range). The new ESD design provides a reduction in seismic noise coupling of about a factor of 100 compared to the conventional design.

\begin{figure}[htbp]
\centering
\includegraphics [width=0.75\textwidth] {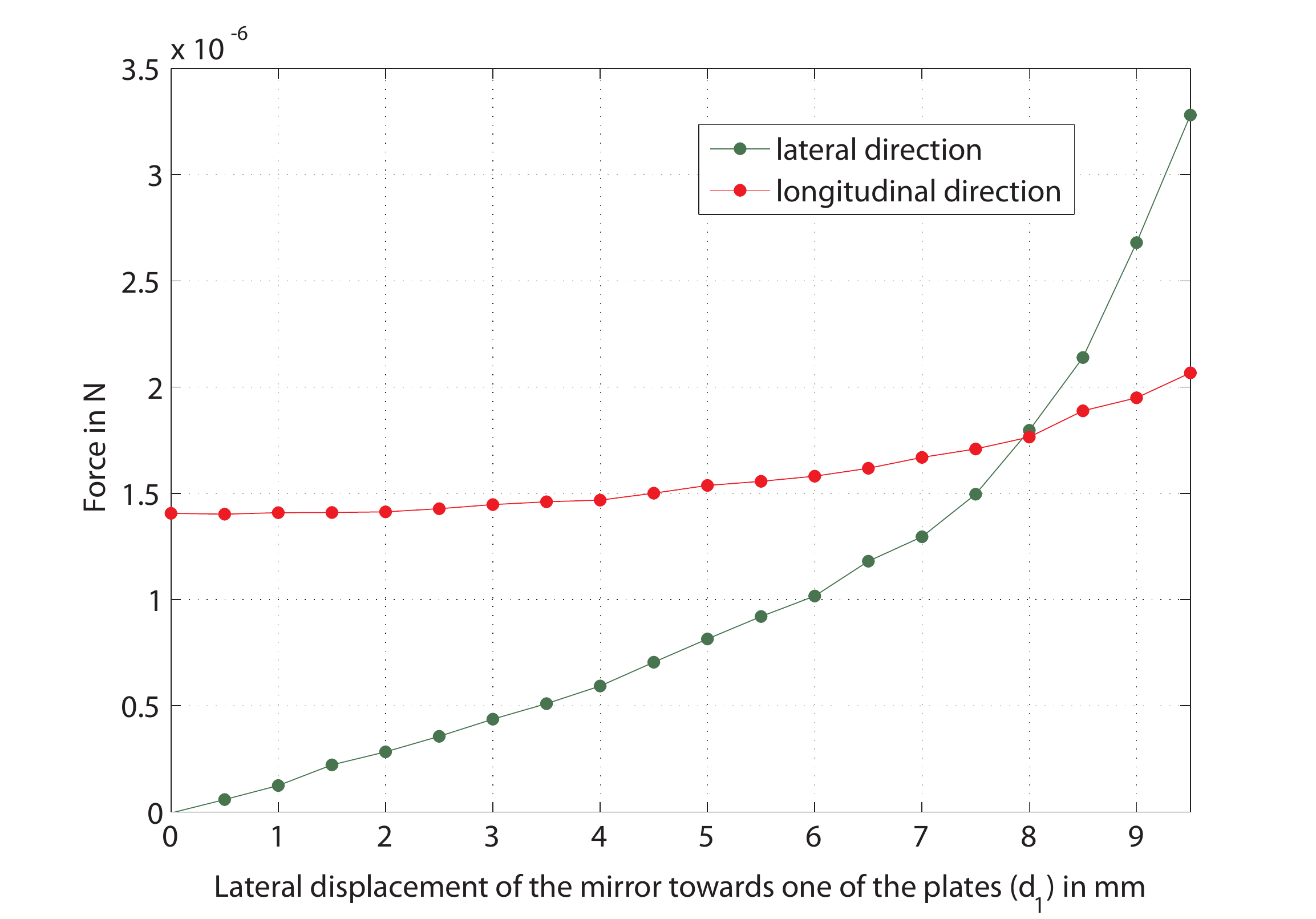}
\caption{Lateral force on the mirror, versus relative lateral position of mirror and ESD.}
\label{fig:force-pos-lat}
\end{figure}

%
%
Moreover, we also investigated how precisely the ESD plates (of the new design) have to be positioned in the direction along their surface normal. If the mirror is closer to one of the ESD plates, it will see a force towards the closer plate. This situation has been simulated using the FE model. The results can be seen in Figure \ref{fig:force-pos-lat}. If we arbitrarily set a limit for the force towards the closer plate to 1/10 of the force in beam direction, then the mirror must be centered between the plates within 1\,mm. 

\section{The Effect of Asymmetries in Materials in the Vicinity of the ESDs}
As mentioned earlier, the main optics of the AEI-10m  interferometer will be hung as a multi-stage pendulum, which is usually the case for gravitational wave detectors. In order to keep the mechanical losses low, the pendulum wires are typically dimensioned such that they are loaded to a significant amount of their breaking stress~\cite{fiber}. In the case of the AEI-10m prototype, it is planned to use fused silica fibers with a diameter of 20\,$\mu$m for the lowest pendulum stage. To protect the main optics if a fiber breaks, a 'catcher' will be placed underneath each suspended mirror.
One could argue that such catchers may alter the performance of the ESDs proposed in this article, as they break the symmetry of the electrostatic environment close to the ESDs. Such asymmetries may introduce a (position-dependent) torque on the mirror, which would lead to coupling of longitudinal mirror motion to mirror alignment. We do not expect the same effect from symmetrically positioned parts around the ESDs. 

\begin{figure}
	\centering
		\includegraphics[width=0.75\textwidth]{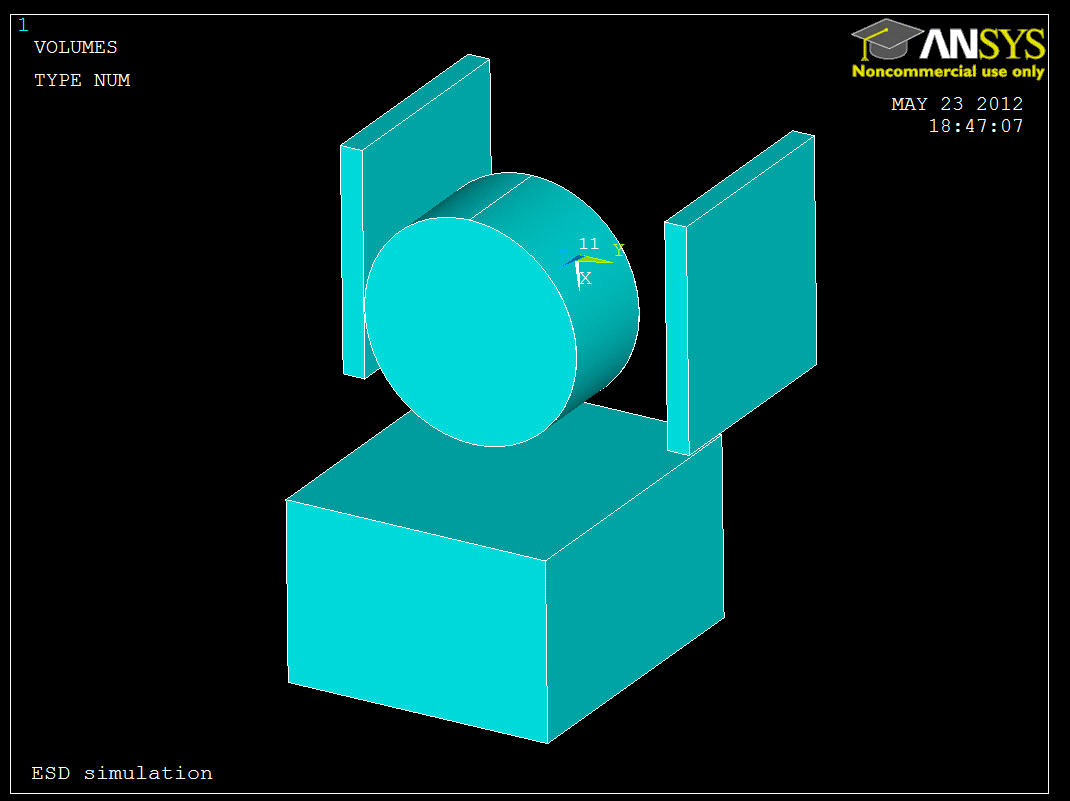}
	\caption{Geometric setup with catcher.}
	\label{fig:catcher}
\end{figure}

We simulated the effect of a sample configuration with catcher, which is modeled as a cuboid, the geometry of which is shown in Figure~\ref{fig:catcher}. In this simulation the ESD plates are at the sides of the mirror. The catcher is positioned in such a way that it is centered below the mirror at the ideal operating point. It sits 1\,cm below the mirror and is 4\,cm tall. In the beam direction, the mirror projects from the ESD by 1\,cm on each side. the catcher is 6.5\,cm wide and is assumed to be at ground potential, while the ESD plates were set to +500\,V and -500\,V.
  \begin{figure}[!ht]
    \subfloat[Force\label{fig:catcher_force}]{%
      \includegraphics[width=0.47\textwidth]{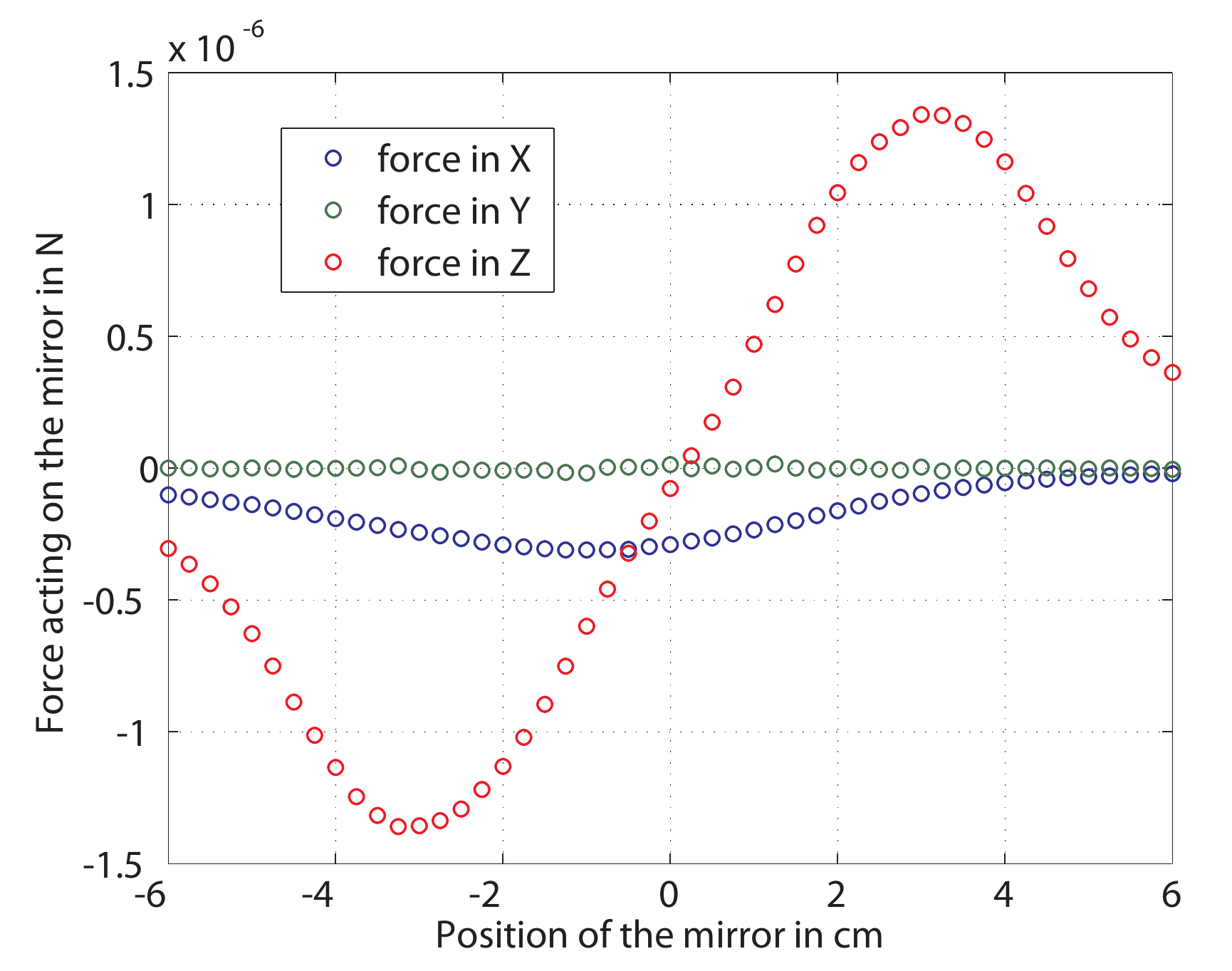}
    }
    \hfill
    \subfloat[Torque\label{fig:catcher_torque}]{%
      \includegraphics[width=0.47\textwidth]{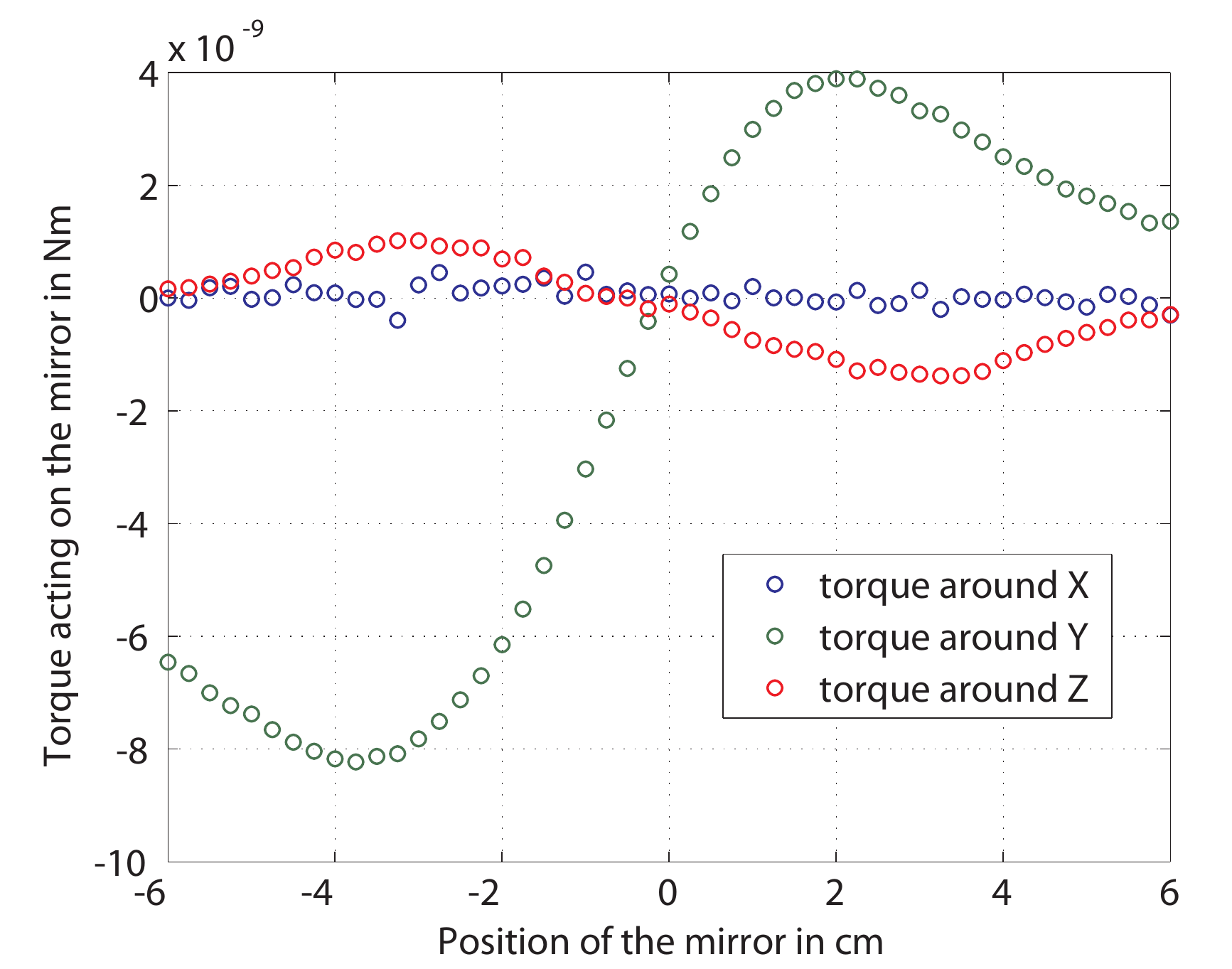}
    }
   	\caption{Force and torque with catcher. The catcher's position was kept constant (with the z-coordinate of  its center of mass at -3.075 cm), while the mirror was moved.}
  \end{figure}
The inclusion of a metal catcher in the simulation caused no significant change in the longitudinal force of the ESD. It did however produce a lateral force of about 20\% of the longitudinal force towards the catcher (at the operating position). Also there will be a torque in the 'tilt' degree of freedom. Fortunately, the torque in Y-direction, which is the strongest, is flat to first order when the catcher is centered with respect to the mirror. It is important to note that this crude catcher geometry is used as a worst case scenario to set an upper limit. To reduce the influence of mechanics around the mirror, one may, for example, use a different catcher geometry or material, such as Macor or polyether ether ketone (PEEK), a vacuum compatible polymer. 
The AEI-10m interferometer suspension will feature 'fibre guards', an aluminium semi-enclosure around the fibres that suspend the mirror. Their presence will necessitate the ESDs to be installed at the top and bottom of the mirrors. 
A detailed description of the design of the 10m prototype suspensions can be found in \cite{giles}.

\section{Summary and Outlook}
We presented a novel type of electrostatic drives for longitudinal test mass actuation in gravitational wave detectors. The ESDs would be plates installed at the sides of the test masses. For the AEI-10m prototype, this type of ESDs could be mounted directly to the seismically isolated table. We find that the ESDs would have to be positioned with an accuracy of better than 3.75\,mm longitudinally and 1\,mm in lateral direction. They could move the mirror by more than 0.28\,$\mu$m with 1\,kV and by more than 1\,$\mu$m when operated at 2\,kV. One aspect that has not been investigated so far refers to the noise terms originating from electrical charges on the test masses. While it would be straightforward to include these in our FE analysis, it is not clear at all what amount of charge and geometrical charge distribution would be sensible to assume. Therefore this aspect needs to be tested experimentally. This work is now underway.

\ack{We thank Peter Fritschel for useful
comments on the manuscript. The authors would like to 
acknowledge the support from
the Max Planck Society, the European Research Council
(ERC-2012-StG: 307245) and the Science and
Technology Facilities Council (STFC, ST/L000946/1). This work was
performed as part of our International Max Planck
Partnership (IMPP) with the Max Planck Society, supported
by SFC, EPSRC and STFC.}

\end{document}